\documentstyle[11pt,moriond,epsfig]{article}

\def\Journal#1#2#3#4{{#1} {\bf #2}, #3 (#4)}

\def\PLB{{\em Phys. Lett.}  B}

\def\NPA{{\em Nucl. Phys.} A}

\def\be{\begin{equation}}
\def\ee{\end{equation}}
\def\bea{\begin{eqnarray}}
\def\eea{\end{eqnarray}}

\newcommand{\jpsi}{J/$\psi$}

\begin{document}
\vspace*{4cm}
\title{PROMPT DIMUONS AND D MESON PRODUCTION IN HEAVY-ION COLLISIONS AT THE SPS}

\author{ B. LENKEIT for the NA60 collaboration }

\address{CERN-EP}

\maketitle
\abstracts{NA60, a follow-up of NA38/50 at the CERN-SPS,
            is a third generation heavy ion experiment finally approved in November
                     2000 for heavy-ion runs in 2002 and 2003. This article will report about the main motivations
                     which lead to this experiment, the main detector concept and the
                     foreseen physics performance.}

\section{Physics motivation}
A very large amount of experimental results were obtained by the CERN SPS
experiments since 1986. A tentative summary has been proposed ~\cite{cernpr}, basically saying
that ``the combined results provide compelling evidence for the
existence of a new state of matter, featuring many of the
characteristics of the primordial soup in which quarks and gluons
existed before they clumped together as the universe cooled down''. Although
this evidence is accepted by a large fraction of the Heavy-Ion community, several
questions remain unclear. Some of these questions will be addressed by NA60,
including a better understanding of the \jpsi\ suppression, by running different 
collision systems like In-In and Pb-Pb, as well as the measurement of the $\chi_c$ 
production in p-N collisions. This article will concentrate on the 
main motivations of the experiment, namely the clarification of the intermediate-
and low-mass dilepton enhancement. An overview of all physics motivations can be found in
Ref. 2 and 3.

\subsection{Intermediate mass dimuons}

The NA38 and NA50 experiments have studied the production of dileptons
in the mass window between the $\phi$ and the \jpsi\ peaks, as a
superposition of Drell-Yan dimuons and simultaneous semileptonic
decays of $D$ and $\bar{D}$ mesons, after subtraction of the
combinatorial background from pion and kaon decays~\cite{qm01-lc}.
The dimuon mass spectra measured in p-A collisions are very well
reproduced taking the high mass region to normalize the Drell-Yan
component and an open charm cross-section in good agreement with
direct measurements made by other experiments.  On the contrary, the
superposition of Drell-Yan and open charm contributions, with the
nucleon-nucleon absolute cross sections scaled with the product of the
mass numbers of the projectile and target nuclei (as expected for hard
processes), fails to properly describe the dimuon yield measured in
ion collisions.

The data can be properly reproduced by simply increasing the open
charm yield, with a scaling factor that grows linearly with the number
of nucleons participating in the collision, reaching a factor 3 in the
most central Pb-Pb collisions.  The observed excess can also be due to
the production of thermal dimuons, a signal that was the original
motivation for the NA38 experiment.  In particular, the intermediate
mass dimuons produced in the most central Pb-Pb collisions are well
reproduced by adding thermal radiation, calculated according to the
model of Ref. 5, to the Drell-Yan and charm contributions
normally extrapolated from nucleon-nucleon collisions.  This model
explicitly includes a QGP phase transition with a critical temperature
of 175~MeV.  The best description of the data is obtained using
$\sim$\,250~MeV as the initial temperature of the QGP medium radiating
the virtual photons.  

The presently available data cannot distinguish
between an absolute enhancement of charm production and the emission
of thermal dilepton radiation.  The clarification of the nature of the
physics process behind the observed excess is one of the remaining questions
of the CERN heavy-ion program and is a basic motivation of NA60.

\subsection{Vector meson resonances}
The CERES experiment has observed that the yield of low mass $e^+e^-$
pairs measured in p-Be and p-Au collisions is properly described by
the expected ``cocktail'' of hadronic decays, while in \mbox{Pb-Au}
collisions, on the contrary, the measured yield, in the mass region
0.2--0.7~GeV, exceeds by a factor 2.5 the expected
signal~\cite{qm99-na45}.  These observations are consistent with the
expectation that the properties of vector mesons should change when
produced in dense matter.  In particular, near the phase transition to
the quark-gluon phase, chiral symmetry should be partially restored,
making the vector mesons indistinguishable from their chiral partners,
thereby inducing changes in their masses and decay widths.  The short
lifetime of the $\rho$ meson, shorter than the expected lifetime of
the dense system produced in the SPS heavy ion collisions, makes it a
sensitive probe of medium effects and, in particular, of chiral
symmetry restoration.

For a final proof of this explanation, high resolution data, especially 
of the low mass resonances, are needed. Besides a good mass resolution better 
statistics are needed compared to the old CERES measurement.
NA60 has the very effective dimuon trigger
inherited from NA50. This will allow to study the low mass resonances with
much higher statistics than CERES.

\section{Experimental apparatus  }

The NA50 experiment \cite{NA50} has been using CERN's highest intensity heavy ion
beam (more than $10^7$~ions per second) and has a very selective
dimuon trigger, quite appropriate to look for rare processes.  The
NA60 experiment complements the muon
spectrometer and zero degree calorimeter already used in NA50 with two
state-of-the-art silicon detectors, placed in the target region: a
radiation hard beam tracker, consisting of four silicon microstrip
detectors placed on the beam and operated at a temperature of 130~K,
and a 10-plane silicon pixel tracking telescope, made with radiation
tolerant readout pixel chips, placed in a 2.5~T dipole magnetic field.
Because of a delay in the delivery of the pixel readout chips, NA60 will do the
proton runs
with a telescope build with silicon strip detectors. This solution only works in
proton induced collisions where the charged particle multiplicity is low enough.

With these two detector systems the interaction vertex can be reconstructed
with an accuracy of roughly 10 $\mu$m (depending on the centrality of the collision).
With the additional magnetic field in front of the absorber, an independent 
measurement of the momentum is available.
Together with the knowledge of the angles, this allows us to match 
the particles before and after
the absorber, the most crucial part in the track reconstruction.
The vertex telescope allows to measure the vertex offset of single muon tracks,
which is important to select muons from D, K and $\pi$ decays. It also improves the mass
resolution of the muon pairs, since the opening angle can be measured before being distorted by the
multiple scattering on the way through the absorber. It also allows to reduce slightly 
the thickness of the muon absorber to improve the acceptance of low momentum muons.

\section{Physics performance capability}

The physics performance of the new experiment was studied by implementing the whole setup
within Geant. The hard probes were generated with PYTHIA, while the soft part was
generated by the Genesis generator previously used by NA50 and CERES. For the underlying background
the VENUS generator was used. 

The simulated data are reconstructed by taking for the
muon spectrometer the standard NA50 reconstruction algorithms. For events with
two reconstructed muon tracks also the target telescope is analyzed starting with  
the beam tracker, which gives the transverse coordinates of the interaction
point. With this information the event vertex can be reconstructed with the help
of all reconstructed tracks in the telescope. Finally the muon tracks can be matched through the absorber and their
impact
parameter, i.e.\ the minimum distance between the
track and the collision vertex, in the transverse plane, can be measured with good accuracy.
  
\begin{figure}[h]
\centering
\begin{tabular}{cc}
\resizebox{0.48\textwidth}{!}{%
\includegraphics*[bb= 41 33 598 598]{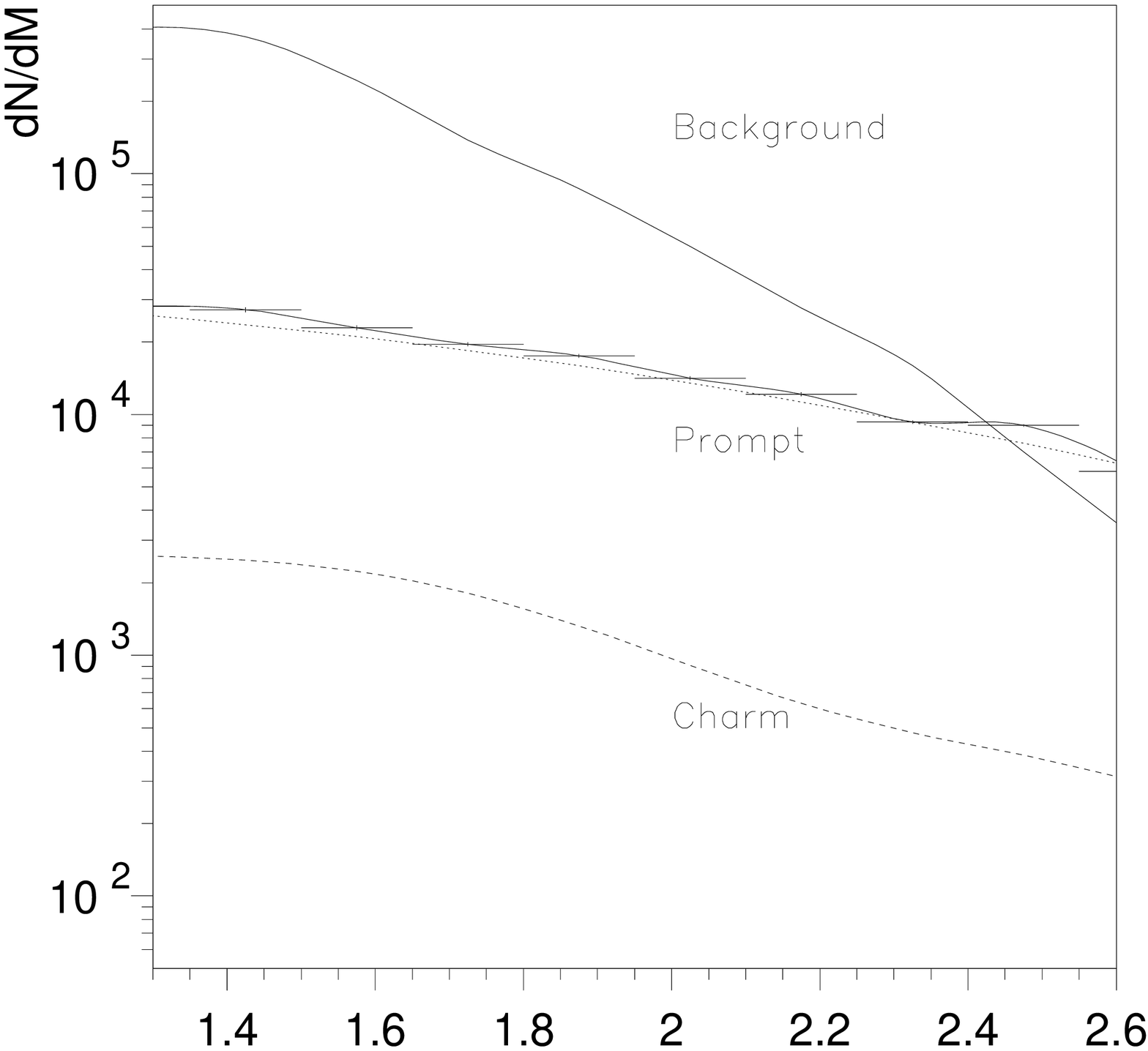}}
&
\resizebox{0.48\textwidth}{!}{%
\includegraphics*[bb= 41 33 598 598]{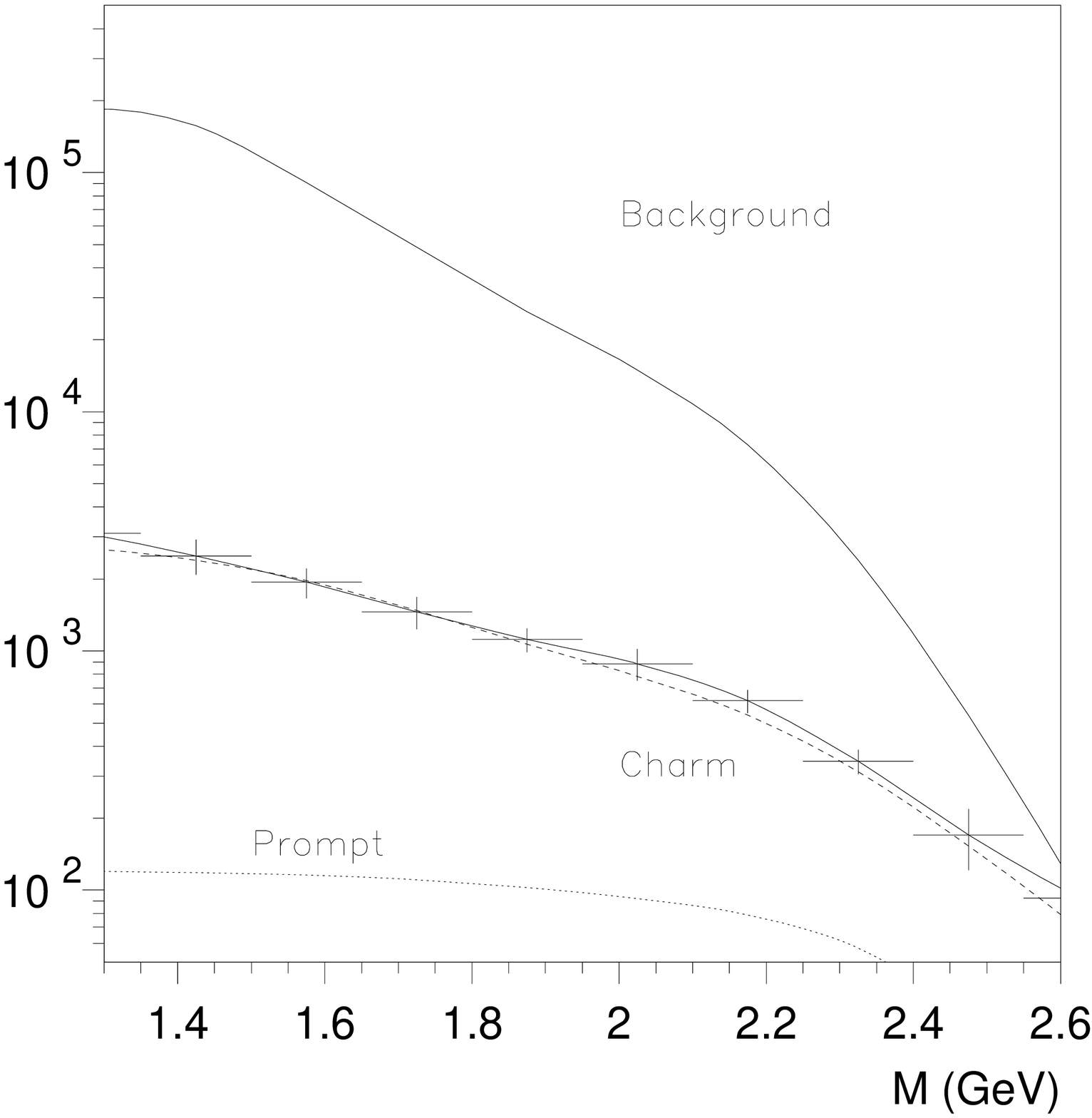}}
\end{tabular}
\caption{Simulated dimuon mass distributions for the prompt (left) and charm
  (right) event selections. The background contribution is also shown,
  including pion/kaon decays and fake matches between the tracks in
  the muon and in the vertex spectrometers.  The error bars in the
  signal points include the uncertainty from background subtraction.}
\label{imr}
\end{figure}

Thanks to
this information, NA60 will be able to separately study the production
of prompt dimuons and the production of muons originating from the
decay of charmed mesons, in p-A and heavy ion collisions.  The prompt
dimuon analysis will use events where both muons come from (very close
to) the interaction vertex.  The open charm event sample is composed
of those events where both muon tracks have a certain minimum offset
with respect to the interaction point and a minimal distance between
themselves at $z_{\mathrm{vertex}}$.  It should not be difficult to see
which of these two event samples is enhanced by a factor of 2 or 3 in
nuclear collisions of $N_{\mathrm{part}}\sim 300$. The result of such a simulated analysis in
the intermediate mass region can be seen in figure \ref {imr}.\hfill\break

% Acceptance picture - discussion

\begin{figure}[ht]
\centering
\begin{tabular}{cc}
\resizebox{0.48\textwidth}{!}{%
\includegraphics*{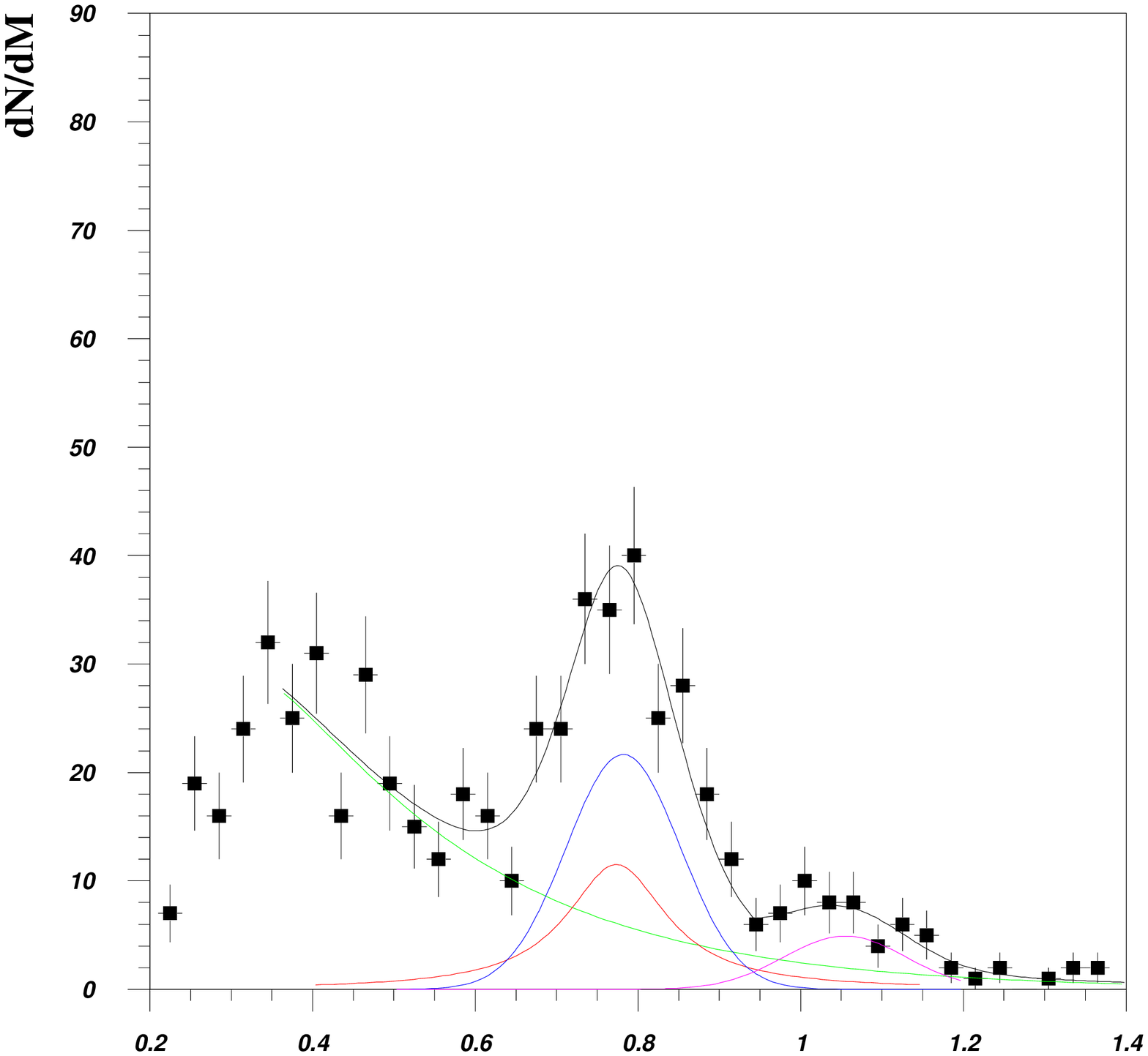}}
&
\resizebox{0.48\textwidth}{!}{%
\includegraphics*{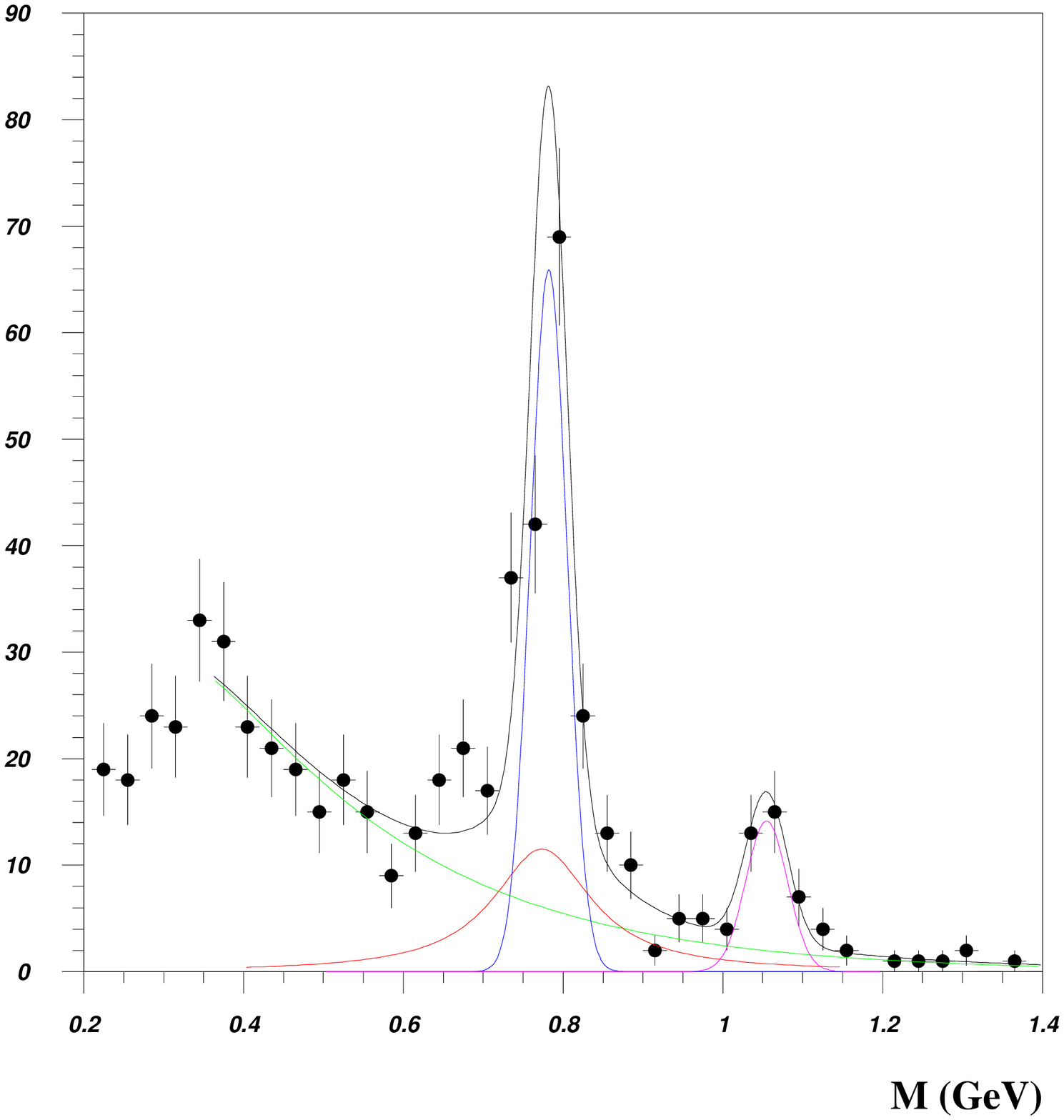}}
\end{tabular}
\vglue-0.4cm
\caption{Dimuon mass distributions measured in 1998, in p-Be
collisions, before (left) and after (right) using the information of
the test pixel telescope.  The curves represent the low mass vector
meson resonances ($\rho$, $\omega$ and $\phi$) on the top of a
continuum.  They are normalized to the same number of events in both
figures. The expected improvement of the mass resolution for the $\omega$ resonance from 70 to 20 MeV
is experimentally confirmed.}
\label{pbe}
\end{figure}

Besides the simulation, the performance in the low mass region was also tested experimentally
with a proton beam.
 The results of this data analysis can be seen in
figure \ref {pbe}. In particular, the improvement in the mass resolution from 70 to 20
MeV at the $\omega$ resonance is experimentally confirmed.

% \subsection{Vector meson resonances}

\section{Summary}

The re-birth of the heavy ion physics program at the CERN SPS, with
the extension of NA49 and the approval of the new NA60 experiment,
represents an evolution from a broad physics program to a dedicated
study of specific signals that already provided very interesting
results. 
The new measurements should give a significant contribution to
the understanding of the presently existing results, and considerably
help in building a convincing logical case that establishes beyond
reasonable doubt the formation (or not) of a deconfined state of
matter in heavy ion collisions at the SPS.

%\section*{Appendix}

\end{document}